\documentclass[a4paper,11pt]{article}
\pdfoutput=1 % if your are submitting a pdflatex (i.e. if you have
\usepackage{jinstpub} % for details on the use of the package, please
\usepackage{lineno}
\title{\boldmath Ultra low background Micromegas detectors for BabyIAXO solar axion search}

\author[a]{E.~Ferrer-Ribas,\note{Corresponding author.}}
\author[b]{K.~Altenmüller,}
\author[a]{B.~Biasuzzi,}
\author[b]{J.F.~Castel,}
\author[b]{S.~Cebrián,}
\author[b]{T.~Dafni,}
\author[c]{K.~Desch,}
\author[b]{D.~Díez-Ibañez,}
\author[b]{J.~Galán,}
\author[b]{J.~Galindo,}
\author[b]{J.A.~García,}
\author[a]{A.~Giganon,}
\author[a]{C.~Goblin,}
\author[b]{I.G.~Irastorza,}
\author[c]{J.~Kaminski,}
\author[b]{G.~Luzón,}
\author[b]{C.~Margalejo,}
\author[b]{H.~Mirallas,}
\author[a]{X.F.~Navick,}
\author[b]{L.~Obis,}
\author[b]{A.~Ortiz de Solórzano,}
\author[c]{J.~von Oy,}
\author[a]{T.~Papaevangelou,}
\author[b]{O.~Pérez,}
\author[d]{E.~Picatoste,}
\author[b,e]{J.~Ruz,}
\author[c]{T.~Schiffer,}
\author[c]{S.~Schmidt,}
\author[a]{L.~Segui,}
\author[b,e]{and J.K.~Vogel}

\affiliation[a]{IRFU, \\CEA, Université Paris-Saclay, Gif-sur-Yvette, France}
\affiliation[b]{Center for Astroparticles and High Energy Physics (CAPA), \\Universidad de Zaragoza, Zaragoza, Spain}
\affiliation[c]{Physikalisches Institut der Universität Bonn, \\ Bonn, Germany}
\affiliation[d]{ICCUB,\\Barcelona, Spain}
\affiliation[e]{ LLNL Lawrence Livermore National Laboratory, \\Livermore, USA}
\emailAdd{esther.ferrer-ribas@cea.fr}

\abstract{The International AXion Observatory (IAXO) is a large scale axion helioscope that will look for axions and axion-like particles produced in the Sun with unprecedented sensitivity. BabyIAXO is an  intermediate experimental stage that will be hosted at DESY (Germany) and that will test all IAXO subsystems serving as a prototype for IAXO but at the same time as a fully-fledged helioscope with potential for discovery. 

One of the crucial components of the project is the ultra-low background X-ray detectors that will image the X-ray photons produced by axion conversion in the experiment. The baseline detection technology for this purpose are Micromegas (Microbulk) detectors. We will show the quest and the strategy to attain the very challenging levels of background targeted for BabyIAXO that need a multi-approach strategy coming from ground measurements, screening campaigns of components of the detector, underground measurements, background models, in-situ background measurements as well as powerful rejection algorithms. First results from the commissioning of the BabyIAXO prototype will be shown.
}

\keywords{Rare event detection, axion, Micromegas}

\arxivnumber{xxxx.xxxx} % only if you have one

\proceeding{7$^{\text{th}}$ International Conference on
Micro Pattern Gaseous
Detectors \\
  11-16 December 2022\\
  Weizmann Institute of Science, Rehovot, Israel}

\newcommand{\gagamma}{g_{a\gamma}}
\newcommand{\ckcs}{counts~keV$^{-1}$~cm$^{-2}$~s$^{-1}$ }
\newcommand{\ma}{m_{a}}
\begin{document}
\maketitle
\flushbottom

\section{Introduction}
Axions are predicted by the Peccei–Quinn mechanism proposed to solve the long-standing strong-CP problem in the Standard Model (SM) of particle physics. More generic Axion-Like Particles (ALPs) appear in diverse extensions of the SM and are invoked in a number of cosmological and astrophysical scenarios. In addition, axions could compose all or part of the cold Dark Matter (DM) and a number of  astrophysical anomalies could also be solved by the presence of axions or ALPs~\cite{DiLuzio:2021ysg}. At present, a large number of experiments is searching for these particles~\cite{Sikivie:2020zpn}. Depending on the source of axions, axions search experiments are classified in laboratory, solar and haloscopes experiments. In this paper we will focus on axion helioscopes, looking for solar axions, that represent the only approach that combines relative immunity to model assumptions  and a competitive sensitivity to parameters that are largely complementary to those accessible by other axion searches.

The CERN Axion Solar Telescope (CAST) is the most sensitive helioscope today. CAST has probed some QCD axion models in the 0.1–1\,eV mass range. The latest CAST result sets an upper bound on the axion-photon coupling, $\gagamma$,
of $<0.66\times10^{-10}$\,GeV up to $\ma\sim0.02$\,eV~\cite{CASTfinal} competing with the strongest bound coming from astrophysics.

The International AXion Observatory (IAXO) is a new-generation large-scale axion helioscope that aims to improve the signal-to-background ratio at four to five orders of magnitude better than CAST~\cite{IAXOPhysics}. This translates into a factor of 20 in terms of the axion–photon coupling constant. This sensitivity improvement relies on the construction of an optimised experiment: a large superconducting eight-coil toroidal magnet with eight X-ray optics focusing the signal photons into small spots that are imaged by low background X-ray detectors. The BabyIAXO experiment, approved at DESY, is the first experimental stage towards IAXO.
In section 2, the BabyIAXO detector requirements and the strategy to reach the required background will be presented. Section 3 will describe the on-going developments on the baseline detector technology, the Micromegas detector.

%%%%%%%%%%%%%%%%%%%%%%%%%%%%%%%%%%%%%
\section{BabyIAXO detector requirements and background reduction strategy }
\begin{figure}[htbp]
\centering 
\includegraphics[width=.35\textwidth]{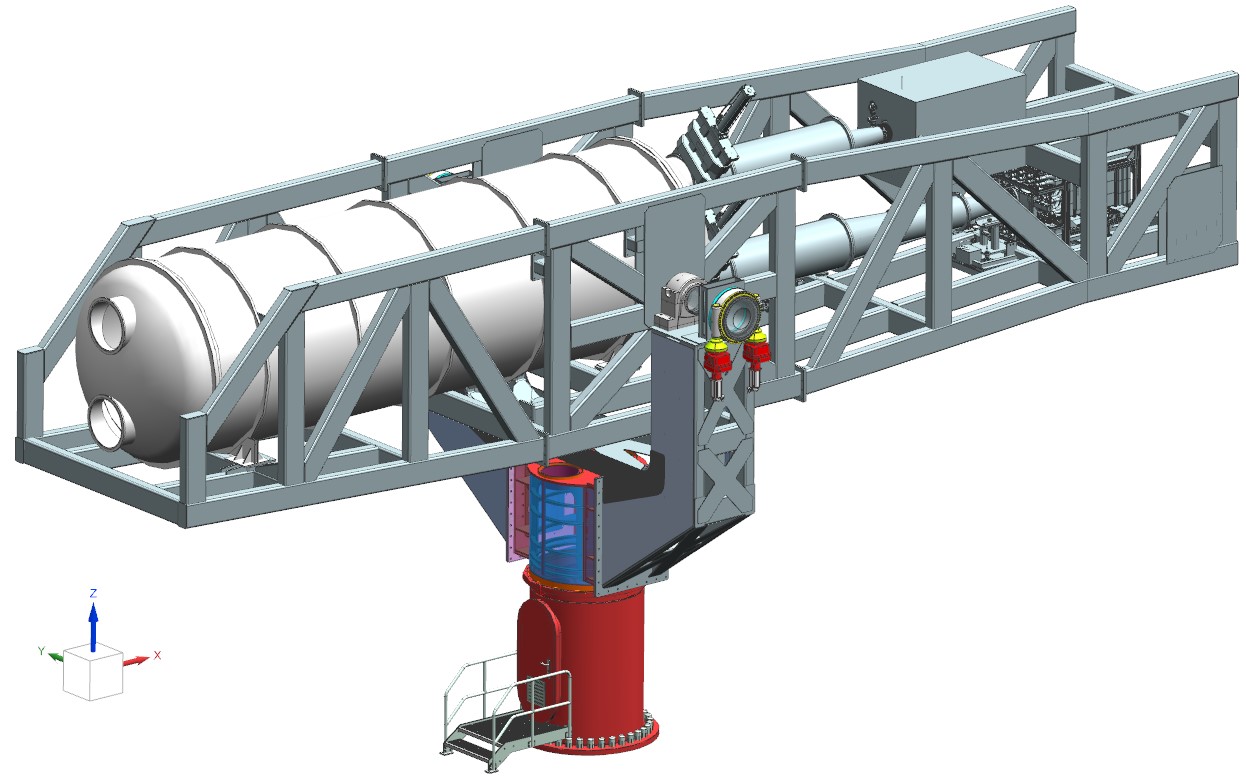}
\caption{\label{fig:BabyIAXO} Overview of the BabyIAXO experiment. The rotating platform, allowing to track the sun for 12 hours per day, sustains the dipole  magnet with two detection lines. Each detection line is composed of X-ray focusing optics and a low background detector.}
\label{ref:BabyIAXO}
\end{figure}

%%%%%%%%%%%%%%%%%%%%%%%%%%%%%%%%%%%%%%
The BabyIAXO experiment consists of a 10\,m long dipole magnet with two 70\,cm diameter bores that constitute two detection lines. Each detection line will be equipped with X-ray focusing optics and a low background detector of similar dimensions and performances to the final ones foreseen for IAXO. The magnet and the detection lines will be placed on a rotating platform that will allow to track the sun for 12 hours per day. The BabyIAXO infrastructure is described in detail~\cite{BabyIAXOCDR}. A sketch of the current design is shown in figure~\ref{fig:BabyIAXO}. The physics reach of the experiment is reviewed in~\cite{IAXOPhysics}. 

The photon energy  of the converted axions via the inverse Primakoff effect in the magnetic field is in the interval 1-10\,keV, with a peak at around 4\,keV~\cite{IAXOPhysics}. The X-ray detectors need to be efficient in this energy range with very low background of the order of 10$^{-7}$\,\ckcs. In order to reach these extremely low values of background in a surface experiment, the detector materials need to be radiopure. Furthermore,  passive and active shielding are essential as well as the use of advanced event discrimination strategies. The baseline detector technology is Time Projection Chambers (TPC) readout by the Micromegas technology~\cite{ReviewMicromegas} based on  the experience and the obtained performance in the CAST experiment. The roadmap to demonstrate target BabyIAXO background levels consists of surface and underground measurements combined with background simulations. The tests at surface will be used to identify background sources and develop methods for reducing the effects. The underground tests,  in the Canfranc Underground laboratory (LSC),  will determine the intrinsic radioactivity (internal or inner shielding components) of the detector. The results of the experimental tests feed background simulations to obtain insight on individual components of the background to support and improve background levels.

Alternative detector technologies are currently under study~\cite{BabyIAXOCDR}: GridPix, Metallic Magnetic Calorimeters (MMC), Neutron Transmutation Doped sensors (NTD), Transition Edge Sensors (TES) and Silicon Drift Detectors (SDD).

\section{On going Micromegas detector developments}
Different protoytypes are being tested at the University of Zaragoza and in Irfu.
On one hand, the IAXO-D0 prototype, used in CAST as a pathfinder for IAXO~\cite{CASTfinal}, is installed in the University of Zaragoza, with a 20\,cm thick lead shielding and with a 4$\pi$ muon veto consisting of three layers of plastic scintillators. Cadmium sheets (1\,mm-thick) are placed between and around the scintillators to capture neutrons. With this strategy the aim is to improve the background by rejecting cosmic muons but also tagging cosmogenic neutrons. Recent results of this prototype~\cite{KonradProceedings} demonstrate a background level of $9\times 10^{-7}$\ckcs.

On the other hand, a new prototype named IAXO-D1,  has been designed and assembled. 
The IAXO-D1 detector shares many technical features with the IAXO-D0 prototype but with improved radiopurity. Details of the basic setup, including the detection concept, thin X-ray windows, readout pattern, gas system and front-end electronics are described in detail in~\cite{BabyIAXOCDR}. The main improvements with respect to the IAXO-D0 prototype are an optimized shielding with active veto and new radiopure electronics. The radiopure electronics are based on an evolution of the AGET chips ~\cite{AGET-Feminos}. The cards have been redesigned with a different partition: the front end card (FEC) will be placed as close as possible to the detector to optimise the signal to noise ratio whereas the back end card will be placed further from the active gas volume separated by the lead shielding. The components of the materials of the FEC have gone through  screening campaigns in the LSC in order to select the most radiopure. A first version of the FEC and back end card have been produced and are being tested.

Different IAXO-D1 prototypes have been assembled and are being characterised in the laboratory with AGET electronics in an Argon-5\% Isobutane mixture. An example of the amplitude signals of the triggered strips for an 
$^{55}$Fe source are shown in figure~\ref{fig:FeSpectrum} left. The reconstructed energy spectrum is shown in figure~\ref{fig:FeSpectrum} right. First characterisation tests show adequate results in terms of gain and energy resolution. 
A second IAXO-D1 prototype has been installed in the Underground Laboratory of Canfranc. It includes an adapted replica of the IAXO-D1 shielding without plastic scintillators. These background measurements should allow
the determination of the  intrinsic  radioactivity of the detector and the limiting factors to achieving the target BabyIAXO background.

\begin{figure}[htbp]
\centering 
\includegraphics[width=.49\textwidth]{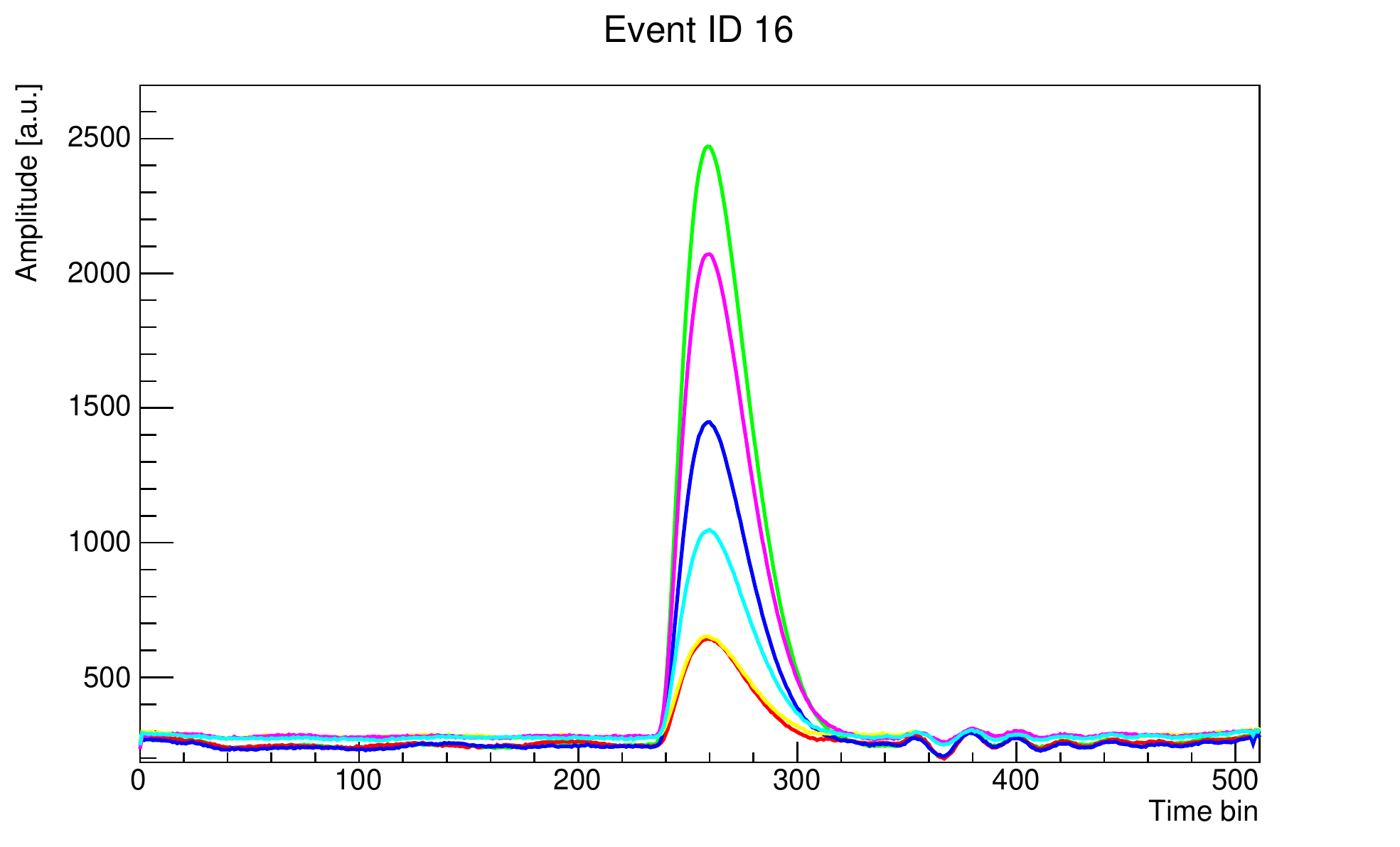}
\includegraphics[width=.49\textwidth]{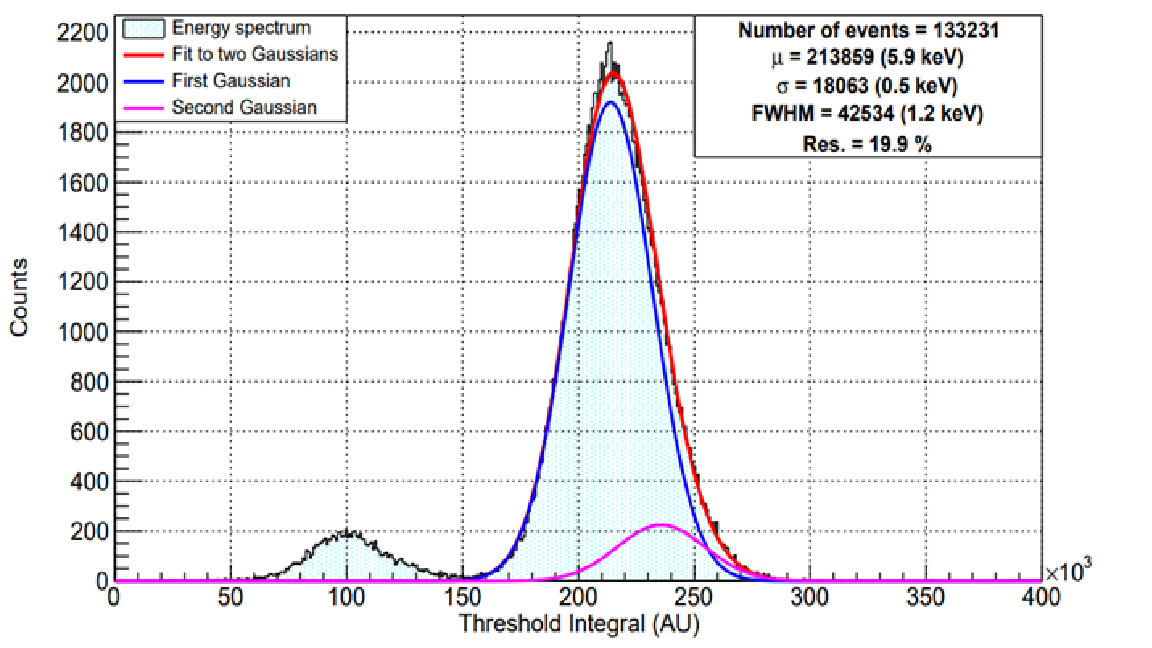}

\caption{\label{fig:FeSpectrum} Left: Typical amplitude signals of the triggered strips for a $^{55}$Fe event. Each triggered strip is represented by a different color. In this particular event, five strips were triggered. Right: Reconstructed energy spectrum for the IAXO-D1 detector in an Argon-5\% Isobutane mixture. Two Gaussians are used to fit the $^{55}$Fe  K$_\alpha$ line and the $^{55}$Fe  K$_\beta$ line of the main peak. The peak at low energy is the Argon escape peak.}
\label{fig:FeSpectrum}
\end{figure}
%%%%%%%%%%%%%%%%%%%%%%%%%%%%%
\section{Conclusion}
%%%%%%%%%%%%%%%%%%%%%%%%%%%%%%%
The BabyIAXO experiment has been approved at DESY (Hamburg) as an intermediate experimental stage of IAXO with relevant physics at reach and with potential for discovery. The construction phase has started and the first systems (platform, optics, detectors) are expected to be commissioned as early as 2024.
The BabyIAXO detector requirements are very challenging in terms of background rejection. Micromegas detectors have been chosen as the baseline technology thanks to their very low background. The IAXO-D1 Micromegas prototype is an optimised prototype made of low radioactive materials with an optimised shielding and a muon external veto. Radiopure electronics are being designed and will be used in the final experiment. Several prototypes have been built and have started operation in Irfu, Zaragoza University and in the underground laboratory of Canfranc with the goal of demonstrating background levels of $\sim 1\times 10^{-7}$\ckcs.
%%%%%%%%%%%%%%%%%%%%%%%%%%%%%%%%%%%%%%%%%%%%%%%%%%%%%%%%%%%%%%%%%

\acknowledgments
We acknowledge support from the Agence Nationale de la Recherche (France) ANR-19-CE31-133 0024. We also acknowledge support from the European Union’s Horizon 2020 research and innovation program under the European Research Council (ERC) grant agreement ERC-2017-AdG788781 (IAXO+) as well as ERC-2018-StG-802836 (AxScale) and under the Marie Skłodowska-Curie grant agreement No 101026819 (LOBRES), and from the Spanish Agencia Estatal de Investigación under grant FPA2016-76978-C3-1-P, the coordinated grant PID2019-108122 GB, and the Maria de Maeztu grant CEX2019-000918-M. Part of this work was performed under the auspices of the U.S. Department of Energy by Lawrence Livermore National Laboratory under Contract DE-AC52-07NA27344.

\end{document}